\documentclass[preprint2]{aastex}

\usepackage{amsmath}
\usepackage{amssymb}
\usepackage{subeqn}

\newcommand{\yr}{{\ensuremath{\mathrm{yr}}}\ }
\newcommand{\km}{{\ensuremath{\mathrm{km}}}\ }
\newcommand{\Msun}{{\ensuremath{\mathrm{M}_{\odot}}}\ }

\newcommand{\Msunyr}{{\ensuremath{\Msun\,\yr^{-1}}}\ }

\newcommand{\lSect}[1]{{\label{sec:#1}}}
\newcommand{\lFig}[1]{{\label{fig:#1}}}
\newcommand{\lEq}[1]{{\label{eq:#1}}}

\newcommand{\pan}[1]{{\textit{#1}}}

\newcommand{\FIGFF}[2]{{\ref{fig:#2}\pan{#1}}}

\newcommand{\Figure}[1]{{Figure~\FIGFF{}{#1}}}

\newcommand{\Sectff}[1]{{\ref{sec:#1}}}
\newcommand{\Sect}[1]{{\S~\Sectff{#1}}}

\newcommand{\Section}[1]{{Section~\Sectff{#1}}}
\newcommand{\Sections}[1]{{Sections~\Sectff{#1}}}
\newcommand{\Eqref}[1]{{\ref{eq:#1}}}
\newcommand{\Eqff}[1]{{(\Eqref{#1})}}

\newcommand{\Equation}[1]{{Equation~\Eqff{#1}}}
\newcommand{\Eq}[1]{{Eq.~\Eqff{#1}}}

\newcommand{\be}{\begin{equation}}
\newcommand{\ee}{\end{equation}}
\newcommand{\ba}{\begin{eqnarray}}
\newcommand{\ea}{\end{eqnarray}}
\newcommand{\f}[2]{\frac{#1}{#2}}
\newcommand{\rhobar}{\bar{\rho}}
\newcommand{\Omegabar}{\mathbf{\bar{\Omega}}}
\newcommand{\Tbar}{\bar{T}}
\newcommand{\sbar}{\bar{s}}
\newcommand{\Pbar}{\bar{P}}
\newcommand{\nubar}{\bar{\nu}}
\newcommand{\kappabar}{\bar{\kappa}}
\newcommand{\Div}{\mathbf{\nabla}\cdot}
\newcommand{\Grad}{\mathbf{\nabla}}

\newcommand{\etal}{\textit{et al.\ }}
\newcommand{\Kepler}{\textsc{kepler}}
\newcommand{\ds}{\delta\hspace{-0.015in}s}
\newcommand{\dP}{\delta\hspace{-0.015in}P}
\newcommand{\dT}{\delta\hspace{-0.015in}T}
\newcommand{\drho}{\delta\hspace{-0.015in}\rho}
\newcommand{\dPhi}{\delta\Phi_g}

\slugcomment{Draft for ApJ, \today} 

\shorttitle{Carbon Ignition in Type Ia Supernovae}
\shortauthors{Kuhlen et al.}

\begin{document}


\title{Carbon Ignition in Type Ia Supernovae: II. A Three-Dimensional
Numerical Model}

\author{M. Kuhlen}

\vskip 0.2 in
\affil{Department of Astronomy and Astrophysics,
University of California, Santa Cruz, CA 95064}
\email{mqk@ucolick.org}

\author{S.~E.\ Woosley}

\vskip 0.2 in
\affil{Department of Astronomy and Astrophysics,
University of California, Santa Cruz, CA 95064}
\email{woosley@ucolick.org}

\and

\author{G.~A.\ Glatzmaier}

\affil{Department of Earth Sciences, University of California, Santa Cruz, CA
95064} 
\email{glatz@es.ucsc.edu}

\begin{abstract}

The thermonuclear runaway that culminates in the explosion of a
Chandrasekhar mass white dwarf as a Type Ia supernova begins centuries
before the star actually explodes. Here, using a 3D anelastic code, we
examine numerically the convective flow during the last minute of that
runaway, a time that is crucial in determining just where and how
often the supernova ignites. We find that the overall convective flow
is dipolar, with the higher temperature fluctuations in an outbound
flow preferentially on one side of the star. Taken at face value, this
suggests an asymmetric ignition that may well persist in the geometry
of the final explosion. However, we also find that even a moderate
amount of rotation tends to fracture this dipole flow, making ignition
over a broader region more likely. Though our calculations lack the
resolution to study the flow at astrophysically relevant Rayleigh
numbers, we also speculate that the observed dipolar flow will become
less organized as the viscosity becomes very small.  Motion within the
dipole flow shows evidence of turbulence, suggesting that only
geometrically large fluctuations ($\sim 1$ km) will persist to ignite
the runaway.  We also examine the probability density function for the
temperature fluctuations, finding evidence for a Gaussian, rather than
exponential distribution, which suggests that ignition sparks may be
strongly spatially clustered.

\end{abstract}


\section{INTRODUCTION}
\lSect{intro}

The leading model for a Type Ia supernova is a carbon-oxygen white
dwarf that accretes matter from a binary companion at a sufficiently
high rate that it is able to grow to the Chandrasekhar mass and
explode \citep[eg.,][]{Nie00}. Much of the research in recent years
has focused upon the computationally challenging problem of how a
nuclear fusion flame, once formed in the white dwarf interior,
succeeds in burning a sufficient fraction of its mass to $^{56}$Ni and
intermediate mass elements to power a healthy explosion and give a
credible light curve and spectrum \citep[eg.,][]{Gam03,Ple04,Roe05a}.
Equally important, however, is just how the flame ignites. A flame
starting from a single point at the middle, a single point off center,
and a multitude of points isotropically distributed around the center
will all give qualitatively different results
\citep{Nie95,Nie96,Ple04,Gar05,Roe05b}. The first numerical simulation
of carbon ignition suggested central ignition \citep{Hof01}, but may
not have correctly represented the pre-explosive convective flow
because of its restricted geometry. Analytic calculations have supported
the idea of off-center ignition, probably at multiple points
\citep{Gar95,Woo03,Wun04}, but require their own assumptions, for
example the persistence of fluctuations and the existence (or non
existence) of an ordered background flow.

The numerical simulation of carbon ignition in this environment
involves a different sort of physics, and optimally, a different sort
of computer code than the flame propagation problem. The fluid motion
is very subsonic (Mach number less than about 0.01) and the density
contrasts that must be followed are small $\sim10^{-5}$. Both can pose
problems for compressible hydrodynamics codes. The simulation must be
followed in 3D for a long time and this is challenging for a code that
is Courant limited by the speed of sound. However, these same
circumstances are very favorable for anelastic hydrodynamics,
which is Courant limited by the fluid velocity. In \Sect{code} we
describe the implementation for the supernova problem of a 3D
anelastic code that has previously been employed to study the sun
\citep{Gla84}, the earth's geodynamo \citep{Gla95}, and planetary
convection \citep{Sun93}. This code is based on spectral methods that
greatly improve the effective resolution over ordinary spatial grids.
As we shall see, resolution is a big issue since the Reynolds number
in the star is Re $\sim 10^{14}$ \citep{Woo03}, far beyond what can be
achieved in any presently existing code. We are thus only able to
glimpse large scale features (\Sect{dipole}) and hints of complex
structures beneath.

Nevertheless, we come to some interesting conclusions. The overall
convective flow, for the range of Reynolds numbers that is accessible,
is dipolar. This has been seen before in different environments
\citep[eg.,][]{Wood03}, but the present study is the first to find it
in numerical simulations of Type Ia supernova ignition. This flow
suggests the supernova may ignite in a lop-sided fashion, but there
are many caveats (\Sect{conclusion}). We are also able to determine
the probability density function (PDF) of the temperature
fluctuations and show that it is Gaussian. This has important
implications for the number of ignition points \citep{Woo03}. We also
call attention to the turbulence that exists in the outgoing ``jet''
of the dipole (\Sect{persistence}). Superimposed on the flow which
carries the sparks that will ignite the supernova is the beginning of
what looks to be a Kolmogorov cascade to smaller scales. This will
limit the size of the temperature fluctuation that is necessary to
ignite the star off center. Sparks that are too small will be dissipated
by turbulence before they run away. It will take a finer resolved
study than the present one to conclusively address the issue, but it
may be that the smallest ignition points are of resolvable size, $\sim
1$ km.

\section{The Initial Model}
\lSect{kepler}

In our implementation of the anelastic approximation we solve for
thermodynamic perturbations to an isentropic, constant reference state
(see \Section{refstate}). The one-dimensional background state has
been determined here using an implicit hydrodynamics code, \Kepler
\citep{Wea78}. A white dwarf composed of 50\% carbon and 50\% oxygen
was allowed to accrete (carbon and oxygen) at a rate of 10$^{-7}$
\Msunyr. When the central density and temperature reached $3.2 \times
10^9$ g cm$^{-3}$ and $2.5 \times 10^8$ K, nuclear energy generation
from carbon fusion exceeded the plasma neutrino loss rate. The excess
energy was carried away by convection. At this point the star's mass
was 1.38 \Msun and its radius 1580 km.

\begin{figure*}[htp]
\includegraphics[width=\textwidth]{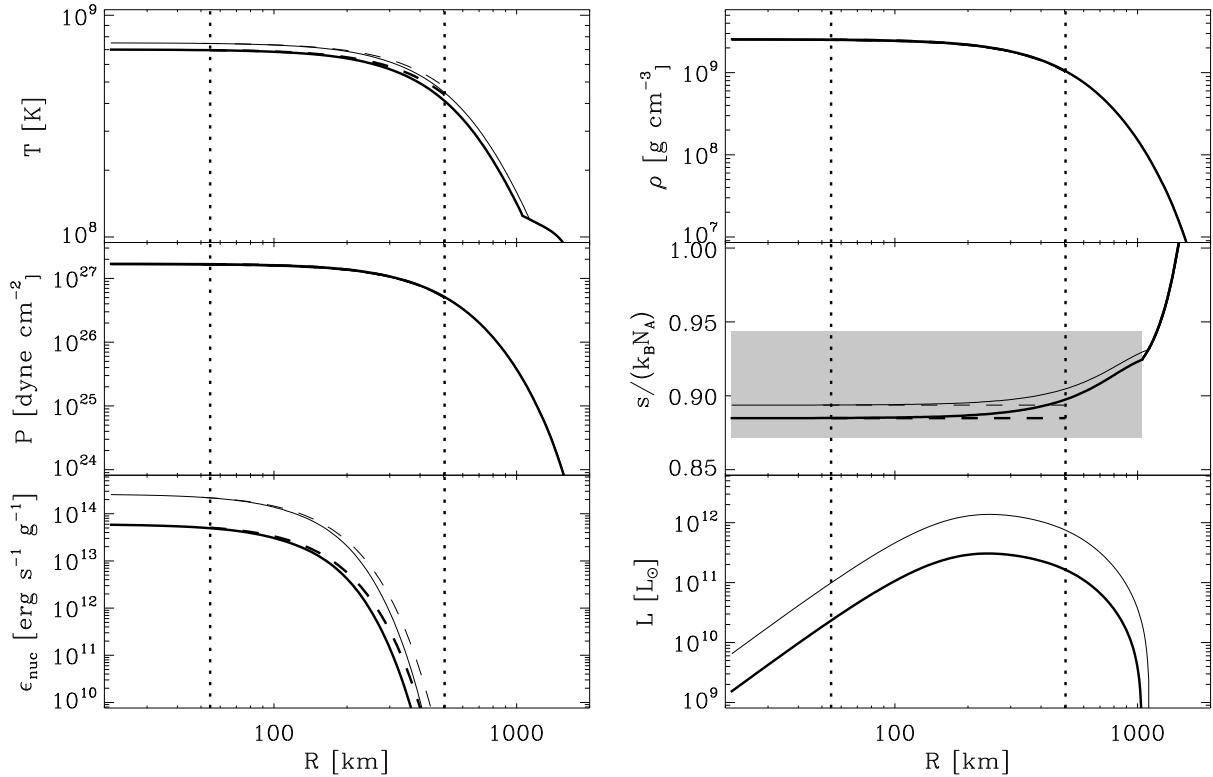}
\caption{
The one-dimensional \Kepler\ model: Temperature, density,
pressure, entropy per baryon, nuclear energy generation, and radiative
luminosity as a function of radius. Thick lines are for the
$T_{c,8}=7$, thin lines for the $T_{c,8}=7.5$ snapshot. The vertical
dotted lines denote the inner and outer boundary of the
three-dimensional anelastic simulation. Dashed lines indicate the
isentropic reference state (see \Section{refstate}). The gray region
in the entropy plot demarcates the range of radii that are
convectively unstable in the \Kepler\ model.
}
\lFig{refstate}
\end{figure*}

Over the next 5000 years the central temperature rose and the extent
of the convective core grew, from initially less than 1\%, to 0.45
\Msun when the central temperature was $4 \times 10^8$ K (2 years
before explosion), and to 0.95 \Msun when the central temperature was
$6 \times 10^8$ K (3 hours before explosion). During this ramp up to
runaway, neutrino losses and radiation transport to the surface of the
star were negligible and a small amount of energy went into
expansion (the central density first increased slightly then declined
by 15\%). Most of the energy went simply into raising the temperature
of the convective core and extending its extent \citep{Bar04}.

When the central temperature reached $7.5 \times 10^8$ K, the
convective core was 1.15 \Msun (1040 km) and the central density, $2.7
\times 10^9$ g cm$^{-3}$. The net binding energy of the white dwarf
(internal plus gravitational) was $4.36 \times 10^{50}$ erg and the
pressure scale height was 450 km. If mixing length convection was left
on, the explosion occurred 190 s later.  Radial profiles of $T$,
$\rho$, $P$, $s m_p/k_B$ (entropy per baryon), $\epsilon_{\rm nuc}$,
and $L$ are shown in \Figure{refstate}, at two different times: when
the central temperature has reached $T_{\rm c,8}=7$ and $T_{\rm
c,8}=7.5$. It is interesting to note that though the entropy in the
convective region in the Kepler model was constant to within 1\%, the
convective flux was not constant for any appreciable span of
radii. Two-thirds of the total nuclear energy ($3.0 \times 10^{45}$
erg s$^{-1}$) was produced within the inner 0.01 \Msun, but the
luminosity itself rose to a maximum of $2.6 \times 10^{45}$ erg
s$^{-1}$ and declined to approximately zero at the outer edge of the
convective core. Almost all of the nuclear energy is lost in transit
to heating along the way.

\section{A THREE DIMENSIONAL MODEL USING ANELASTIC HYDRODYNAMICS}
\lSect{anelastic}

\subsection{The Code and its Modifications}
\lSect{code}

The code used for the simulations presented here is based upon the
anelastic hydrodynamics code of \citet{Gla84}. The hydrodynamic
equations are implemented in a fully spectral manner, expanding all
quantities in spherical harmonics to cover the angular variations, and
in Chebyshev polynomials for the radial dependence. Although a
spectral method in spherical coordinates is more difficult to
implement than one based on a finite differencing scheme on a
Cartesian grid, and also requires more communication on a
multi-processor parallel computation system, a major advantage is the
greater efficiency provided by the spectral method. To achieve a
modest accuracy of a few percent, spectral methods typically require
only half as many degrees of freedom {\em per dimension} as a fourth
order finite difference method \citep{Boy00}. Chebyshev polynomials
are defined on a domain from $-1$ to $+1$, which in this code is
mapped to the inner and outer boundary of the convective
region. Unfortunately, the current implementation of spherical
coordinates prohibits the computational domain from extending all the
way to the origin, since $1/r$ terms in the angular derivatives lead
to divergences. A central sphere is cut out and replaced by an inner
impermeable boundary condition, even in situations where the
convective region includes the center of the star (see
\ref{sec:models}). The code employs a rhomboidal truncation of the
spherical harmonic modes, which means that every longitudinal Fourier
mode ($e^{im\phi}$) has the same number of latitudinal associated
Legendre functions ($P_l^m(\cos \theta)$). A rhomboidal truncation
scheme is easier to parallelize than a more conventional triangular
truncation ($-l \le m \le +l$), and it ensures that the latitudinal
resolution is the same for all longitudinal modes.
 
In this formulation of anelastic hydrodynamics, the mass, momentum,
and entropy conservation equations are expanded in a power series
around a one-dimensional, constant, and isentropic reference state,
and only the lowest orders terms are retained. The reference state is
determined by the one-dimensional stellar evolution code \Kepler\ (see
\Sections{kepler} and \ref{sec:refstate}). In the following all
reference state quantities are barred, and all perturbations denoted
with a $\delta$. The fundamental quantities in this code are the
entropy ($\ds$) and pressure ($\dP$) perturbations. In order to relate
these to the more familiar quantities density and temperature, we
require an equation of state.

\begin{subequations}
\ba
\dT & = & \left(\overline{\f{\partial T}{\partial s}}\right)_P \ds + \left(\overline{\f{\partial T}{\partial P}}\right)_s \dP \\
\drho & = & \left(\overline{\f{\partial \rho}{\partial s}}\right)_P \ds + \left(\overline{\f{\partial \rho}{\partial P}}\right)_s \dP
\ea
\lEq{eos}
\end{subequations}

The equation of state is specified by four thermodynamic partial
derivatives, which are part of the time-independent reference state
and depend only on radius. We use the Helmholtz equation of state code
\cite[]{Tim00} to calculate these derivatives for a given \Kepler\
reference state.

In total, we solve for two thermodynamic variables ($\ds$ and $\dP$),
the perturbation of the gravitational potential ($\dPhi$), and
the three components of the fluid flow velocity ($v_r, v_\theta,$ and
$v_\phi$). The relevant equations are:

\begin{subequations}
\ba
\Div(\rhobar\mathbf{v}) & = & 0\\
\rhobar\f{\partial \mathbf{v}}{\partial t} & = & -\Div(\rhobar\mathbf{v}\otimes\mathbf{v})-\rhobar \Grad(\f{\dP}{\rhobar}+\dPhi) \nonumber \\
& & -\left(\f{\partial \rhobar}{\partial s}\right)_P\ds\;\bar{g}\;\mathbf{\hat{r}} \nonumber + 2\;\rhobar\;\mathbf{v}\times\mathbf{\bar{\Omega}} \\
& & +\Div \left(2\rhobar\nubar(\bar{\mathbf{\mathsf{E}}}
-\f{1}{3}(\Div \mathbf{v})\mathbf{\mathsf{I}})\right)\\
\rhobar \Tbar \f{\partial \ds}{\partial t} & = & \Div(\kappabar\rhobar 
\Tbar \Grad \ds)- \Tbar \Div(\rhobar \ds \mathbf{v})\nonumber \\
& & +\rhobar(\bar{\epsilon}_{\rm nuc}+\delta\epsilon_{\rm nuc}) \nonumber \\
& & -\rhobar(\bar{\epsilon}_{\nu}+\delta\epsilon_{\nu}) \\
\Grad^2\dPhi & = & 4\pi G \drho
\ea
\lEq{anelastic}
\end{subequations}

Here $\bar{\mathbf{\mathsf{E}}}$ is the rate of strain tensor,
$\mathbf{\mathsf{I}}$ the identity matrix, and $\bar{\Omega}$ the
reference state rotation vector. In order to allow for the differential
heating of hot and cold convective elements, we have included a first
order perturbation term in addition to the one-dimensional background
nuclear burning rate. This background energy generation rate is
determined from the \Kepler\ output and fitted to a power law, $\rho
T^\alpha$. For the simulations discussed in this paper $\alpha$ is
approximately equal to 27. The total energy generation rate is then:

\be 
\epsilon_{\rm nuc} = \bar{\epsilon}_{\rm nuc} + \epsilon'_{\rm nuc} \propto \rhobar \Tbar^\alpha \left( 1 + \f{\drho}{\rhobar} + \alpha \f{\dT}{\Tbar} \right)
\ee

Energy losses to neutrinos ($\bar{\epsilon}_{\nu},\epsilon_{\nu}$) act
as a sink in the entropy conservation equation. Results from one
dimensional simulations indicate that these neutrino losses are
negligible (\Section{kepler}), and we have included them in
\Equation{anelastic} only for completeness.

Momentum and energy transport at scales below our spatial resolution
are handled by introducing artificially high viscous and thermal
diffusion coefficients $\nubar$ and $\kappabar$. In our simulation the
Prandtl number (Pr $= \nubar/\kappabar$) is kept at unity, and $\nubar$
and $\kappabar$ are lowered as much as possible for a given resolution
(see \Section{models}). These turbulent diffusion coefficients also
enter in the calculations of the dimensionless numbers characterizing
the fluid flow, the Rayleigh (Ra) and Reynolds (Re) numbers.

\begin{subequations}
\ba
{\rm Ra} & \sim & \f{D^3 g \Delta s}{c_P\nubar\kappabar} \\
{\rm Re} & \sim & \f{v D}{\nubar}
\ea
\end{subequations}

\noindent Here $D$ is the depth of the modeled region and $\Delta
s$ the change in specific entropy across $D$. Ra is proportional to
the ratio of buoyancy to diffusion forces, and Re is proportional to
the ratio of inertial to viscous forces. A large Rayleigh number
indicates very vigorous convection, and a large Reynolds number
indicates a highly turbulent flow. Typical Rayleigh numbers in the
central convective region of a pre-SNIa white dwarf are around
$10^{25}$ and typical Reynolds numbers around $10^{14}$
\cite[]{Woo03}. Use of the turbulent viscous and thermal diffusion
coefficients instead of the true molecular ones greatly reduces the
magnitude of Ra and Re that are achievable with this code. Our
simulations reach Ra $\approx 10^7$ and Re $\approx 1500$, just
entering the turbulent convective regime, but a far cry from the true
physical conditions (see
\Section{results}).

\subsection{Mapping the Background State into the 3D-Code}
\lSect{refstate}


From a given \Kepler\ output (\Section{kepler}) we extract the density
and pressure profiles for the region of interest. For numerical
stability, we then fit a polytrope ($P(r)=K\rho(r)^{1+1/n}$) to this
profile, and solve the resulting Lane-Emden equation for $\bar{P}(r)$
and $\bar{\rho}(r)$. Using the Helmholtz equation of state code
\cite[]{Tim00} we solve for the corresponding temperature profile
$\bar{T}(r)$ and for the thermodynamic partial derivatives (see
\Section{code}). Since the \Kepler\ models are close to, but not
completely isentropic we also determine a volume averaged background
state entropy. This quantity is useful for comparisons with analytical
models, but never used in the 3D anelastic calculation itself.

\subsection{Models and Procedures}
\lSect{models}

Analytical considerations show \cite[]{Woo03} that the ignition of the
final deflagration leading to the supernova explosion is likely to
start somewhere off-center at about $150 - 200$ km. We have chosen to
model a region extending from $50$ km to $500$ km. The boundary
conditions at these radii are impermeable and stress-free ($v_\perp=0$
and $\frac{\partial}{\partial
r}\left(\frac{v_\parallel}{r}\right)=0$).

Since almost all of the energy generation and the most vigorous
convection occur at radii less than $200 \km$, we don't expect this
outer boundary at $500 \km$ to unduly influence our results. The
central boundary is more problematic, since in a real star nothing
prevents fluid from streaming through the center. In our model cold
sinking fluid will splash around the inner boundary. Only by
extrapolating from conditions outside $R=50 \km$, can we say anything
about the flow pattern right at the center.

\begin{table*}[htp]
\begin{center}
\caption{Summary of models}
\vspace*{0.1in}
\begin{tabular}{cccccccl}
\tableline
\tableline
Name & $T_{c,8}$ & $n_{\rm max}$ & $m_{\rm max}$ & $l_{\rm max}$ & $\Delta t$ & $\Omega$  & Description \\
& & & & & [s] & [rad s$^{-1}$] & \\
\tableline
C7 & 7.0 & 241 & 42 & 85 & 70.36 & 0.0 & fiducial model \\
C7rot & 7.0 & \ldots & \ldots & \ldots & 77.06 & 0.167 & rotating background state \\
C75 & 7.5 & \ldots & \ldots & \ldots & 40.90 & 0.0 & higher central temperature \\
\tableline
\end{tabular}
\end{center}
\end{table*}

Additional boundary conditions are necessary for the entropy equation.
Since the radial velocity, and with it, the convective heat flux, is
forced to zero at the inner and outer boundary, turbulent diffusion is
the only mechanism by which energy can be transported across these
boundaries. This heat diffusion is carried by a negative radial
entropy gradient. At the inner boundary the \Kepler\ luminosity is
essentially zero, and so we set $d\ds/dr=0$ there. At the outer
boundary the \Kepler\ luminosity is non-zero ($\sim 10^{45}$ -
$10^{46}$ erg/s), but still much less than the total integrated energy
generation rate in the simulated volume. The excess energy deposited
on the grid gradually leads to an increase in central temperature as
described in \Section{kepler}. Our anelastic model, however, is
constructed around a time-independent reference state. The inability
to follow the evolution of the background state forces us to examine
several ``snapshots'' in the evolution. In this paper we consider two
different central temperatures: $T_{c,8}=$ 7.0 and 7.5. Our goal is to
establish a pseudo-steady-state at each of these epochs, which, while
being physically unrealistic, might still allow us to infer something
about the spectrum of temperature fluctuations and the global flow
pattern. To do this we compensate for the excess energy by adjusting
the outer luminosity to balance the total energy generation rate in
our computational domain. In order to establish a sufficiently large
negative entropy gradient, $\ds$ is forced to become negative at
the outer boundary. This is artificial -- in a real white
dwarf the entropy gradient would remain adiabatic throughout the
convection zone. The compromise was necessary in order to achieve a 
model in near steady state that would run stably for a long time.

\begin{figure}[htp]
\includegraphics[width=\columnwidth]{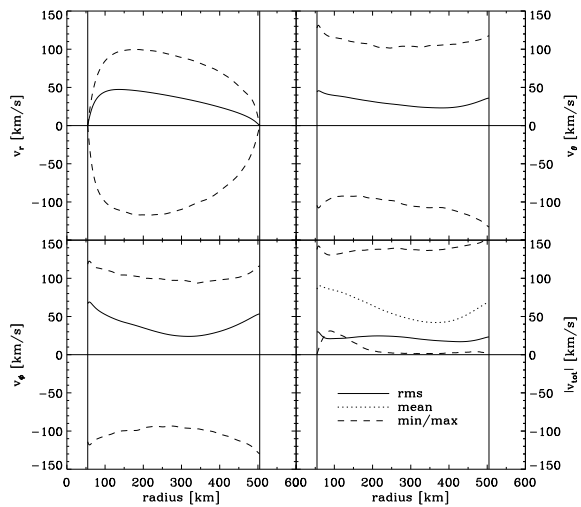}
\caption{
The horizontally and time averaged radial, angular, and total
velocities as a function of radius in the C7 simulation. The solid
line shows the root mean square, the dashed lines the minimum and
maximum, and the dotted line (only in the total velocity plot) the
mean velocities in each radial shell. The radial velocity is forced to
zero at the inner boundary (see text for discussion).
}
\lFig{velocities}
\end{figure}

\begin{figure*}[htp]
\includegraphics[width=\textwidth]{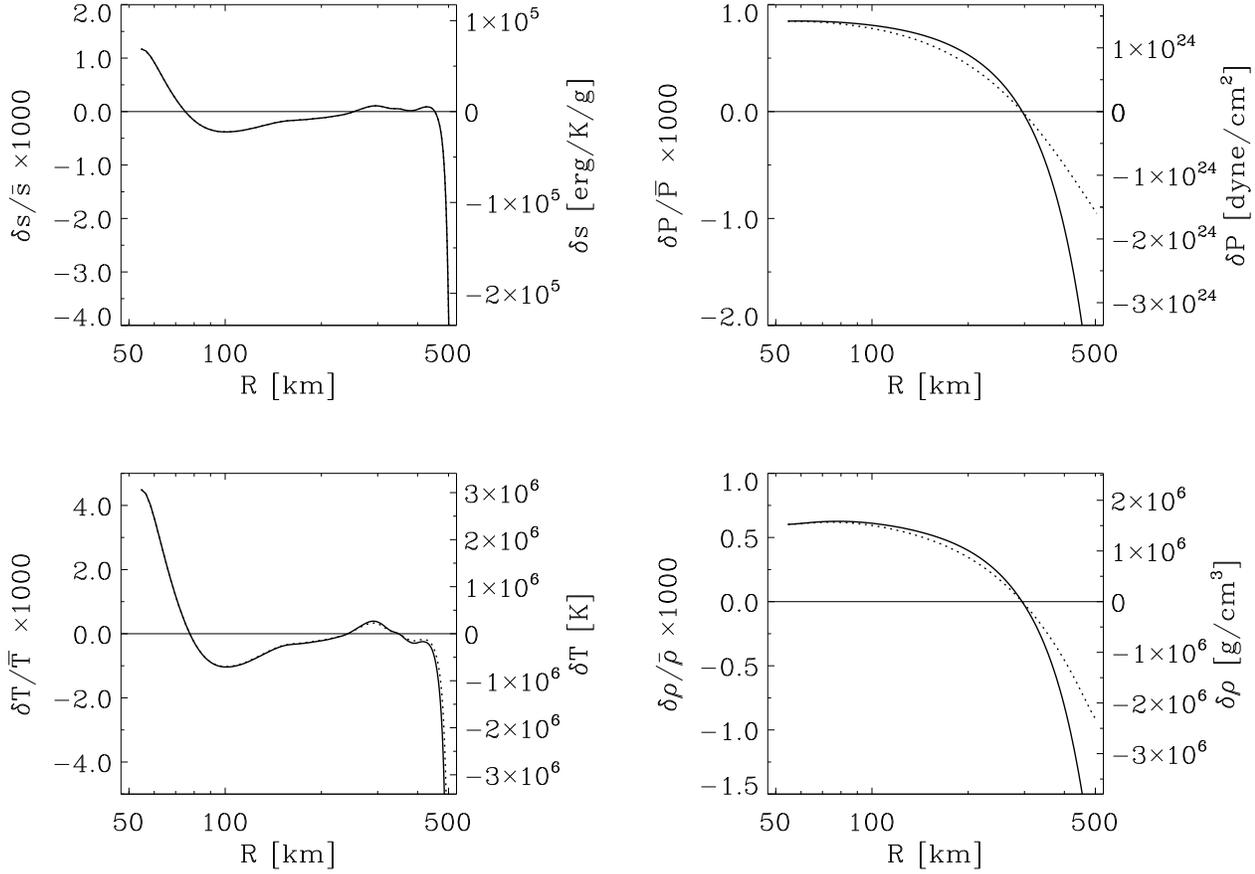}
\caption{ 
Horizontally averaged mass-weighted radial profiles of $\delta
X/\bar{X}$ (solid, left ordinate) and $\delta X$ (dotted, right
ordinate), where $\delta X$ is the entropy, pressure, density, and
temperature perturbation (clockwise from top left).
}
\lFig{profiles}
\end{figure*}

In order to allow the model to find a stable solution quickly, we
begin with large turbulent diffusivities and gradually turn them down,
in small enough increments that the code is able to find a new stable
solution. This decrease raises our operational Rayleigh and Reynolds
numbers. To ensure numerical stability the diffusivities must remain
large enough to prevent a build-up of entropy at the smallest scales.
We continue to lower the diffusivities to the smallest value that can
be accommodated with a given spatial resolution. We then continue to
evolve the model at fixed diffusivity. In this manner we found
pseudo-steady-states for both temperature snapshots. The results
presented here are from only the last $84,000$ time steps, whereas the
relaxation phase of the simulation took about $175,000$ time steps.
The long duration of the relaxation ensures that any ``memory'' of the
initial conditions of the simulation is completely erased.

We considered three models: C7, a non-rotating model with a central
temperature of $T_{\rm c,8}$=7.0, C7rot, identical to C7, except that
the background state is rotated rigidly with an angular velocity
$\Omega = 0.167$ rad/s (Ekman number $\sim 7\times 10^{-4}$, see
\Section{rotation}), and C75, non-rotating, but with $T_{\rm
c,8}$=7.5. The resolution was the same in all three models: $n_{\rm
max}=237$ Chebyshev polynomials in the radial direction, and spherical
harmonics up to $m_{\rm max}=42$ and $l_{\rm max}=85$. As we show
below (\Section{results}), typical plume velocities are $\sim 50$ km/s
and $\sim 100$ km/s in the C7 and C75 simulations, respectively, which
translates to convective turnover times of $t_{\rm to} = 2 (R_{\rm
outer} - R_{\rm inner})/v_{\rm plume} =$ 18 and 9 seconds
respectively. The two simulations are evolved for about 70 and 40
seconds star time, so for roughly four convective turnovers in both
cases. The parameters of each simulation are summarized in Table~1.

\section{RESULTS}
\lSect{results}

\subsection{Velocities and Thermodynamic Perturbations}
\lSect{vel+pert}


In order for the anelastic approximation to be valid, the fluid flow
must remain subsonic and variations from the reference state must be
small, $< \sim 1\%$. The time and angle averaged (over the last 84000
time steps) total, radial, and angular velocities are plotted as a
function of radius in \Figure{velocities}. The horizontal mean as a
function of radius of the absolute value of total velocity increases
from a minimum of about 50 km/s at $R=400$ km up to about 90 km/s near
the center. The maximum velocity achieved anywhere on the grid is 150
km/s. This is much less than the sound speed of $c_{\rm sound}
\sim$8000 km/s, so the condition of subsonic flow is satisfied.

\Figure{profiles} shows horizontally averaged mass-weighted radial
profiles of the thermodynamic perturbations at the end of the C7
simulation. The top two plots show the two independent variables, the
entropy and pressure perturbations. The two perturbations are
comparable in magnitude, and both lie below the 1\% level. The maximum
entropy and pressure perturbations in the whole volume, of course, are
higher: $(|\ds|/\sbar)_{\rm max}=0.015$ and $(|\dP|/\Pbar)_{\rm
max}=0.003$. The largest entropy perturbation is negative, occurs at
the outer boundary, and is caused by the large negative entropy
gradient required to balance the total energy generation, as explained
in \Section{models}. Using \Eq{eos} we can determine temperature and
density fluctuations from the entropy and pressure
perturbations. These are shown in the lower two plots of
\Figure{profiles}. Since $\left(\overline{\f{\partial \ln T}{\partial
\ln s}}\right)_P \gg \left(\overline{\f{\partial \ln T}{\partial \ln
P}}\right)_s$ and $\left(\overline{\f{\partial \ln \rho}{\partial \ln
s}}\right)_P \ll \left(\overline{\f{\partial \ln \rho}{\partial \ln
P}}\right)_s$,
\begin{subequations}
\be
\dT/\Tbar \approx \left(\overline{\f{\partial \ln T}{\partial \ln s}}\right)_P \ds/\sbar,
\ee
\be
\drho/\rhobar \approx \left(\overline{\f{\partial \ln \rho}{\partial \ln P}}\right)_s \dP/\Pbar.
\ee
\end{subequations}

\begin{figure}[htp]
\includegraphics[width=\columnwidth]{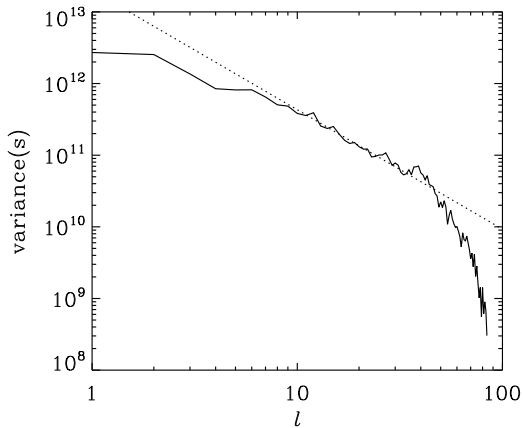}
\caption{ 
Variance of entropy perturbation versus spherical harmonic
wavenumber $l$ at the end of the C7 simulation. The dotted line is a
$l^{-5/3}$ power law, as expected for a Kolmogorov turbulent spectrum.
}
\lFig{kolmogorov}
\end{figure}

\noindent For the conditions inside the simulated region,
$\left(\overline{\f{\partial \ln T}{\partial \ln s}}\right)_{} \approx
3.5$ and $\left(\overline{\f{\partial \ln \rho}{\partial \ln
P}}\right)_s \approx 0.75$. The largest positive temperature and
density perturbation is 1.5\% and 0.7\%, respectively. Since all
thermodynamic perturbations remain below the few percent level, we
conclude that the anelastic approximation is indeed a valid one for
the problem at hand.

Our resolution of $n_{\rm max}=237$, $m_{\rm max}=42$, and $l_{\rm
max}=85$ allowed us to lower the diffusivities to $\kappabar = \nubar = 5
\times 10^{11} {\rm cm}^2/s$, corresponding to Rayleigh and
Reynolds numbers of Ra $= \sim 10^7$ and Re $= \sim 1500$\footnote{We
used $v_{\rm max}$ in the calculation of Re. Using $v_{\rm mean}$
would lower this estimate by about a factor of three.}, in all three
simulations. The transition from laminar to turbulent flow generally
occurs close to Re=2000, so our simulations have not quite reached the
regime of fully turbulent convection, but aren't fully laminar
either. In fact, we see about $2/3$ of an order of magnitude of a
Kolmogorov turbulent cascade in an angular power spectrum of
entropy perturbation, \Figure{kolmogorov}. The turbulent cascade
begins at $l=10$ and extends down to $l=50$.

While the requirements of a time-independent background state and
small perturbations prevent us from following the increase of the
central temperature all the way to the point where localized flames
develop, we can look at the C75 snapshot to get an idea of what the
fluid flow looks like at a higher central temperature of $7.5 \times
10^8$ K. This study employed the same resolution as in the C7
simulation, but with the more energetic background state, we were only
able to lower the diffusivities to $\kappabar = \nubar = 8.5 \times 10^{11}
{\rm cm}^2/s$. Due to the strong temperature dependence of the nuclear
energy generation rate, the convection is much more vigorous in the
C75 simulation. Both the velocities and thermodynamic perturbations
are larger than in the C7 simulation. The mean velocity now ranges
from 70 to 130 km/s, and the maximum velocity on the grid is 276
km/s. These numbers and their temperature scaling agree well with the
analytic expectations \citep{Woo03}.  Typical thermodynamic
perturbations have also increased: the shell averaged temperature
fluctuation reaches 2\% at the innermost radial bin, with the largest
positive perturbation reaching 4.4\%. Higher temperature perturbations
and larger radial velocities increase the likelihood of an off-center
supernova ignition. This possibility is further discussed in
\Section{persistence}.

\subsection{Flow Patterns}
\lSect{dipole}

\begin{figure*}[htp]
\includegraphics[width=\textwidth]{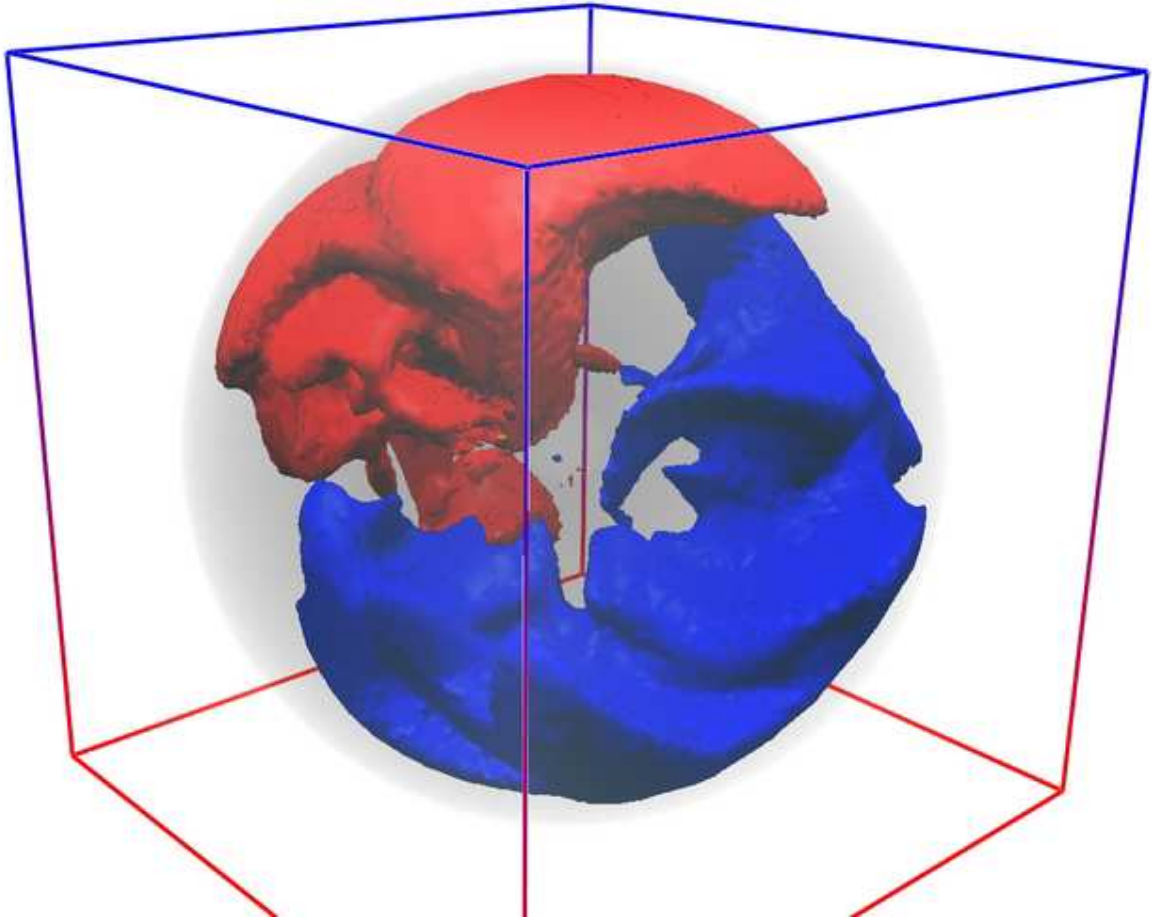}
\caption{
Two iso-velocity surfaces at $|v_r|=40$ km/s. The sharp division
between infalling (red) and outflowing (blue) material clearly
demonstrates the dipole nature of the flow.
\lFig{dipole_flow_3D}
}
\end{figure*}

The most striking feature of our calculation is a large coherent
dipolar circulation. \Figure{dipole_flow_3D} shows a three-dimensional
representation of this flow - two iso-radial-velocity surfaces at $v_r
= \pm 40$ km/s. The outflowing surface (blue) is predominantly located
in one hemisphere, the inflowing surface (red), in the other. The
dipole nature of the flow can also be seen in \Figure{slices}, which
shows two-dimensional equatorial and meridional slices of the
temperature fluctuation and the radial component of velocity.

\begin{figure*}[ht]
\includegraphics[width=\textwidth]{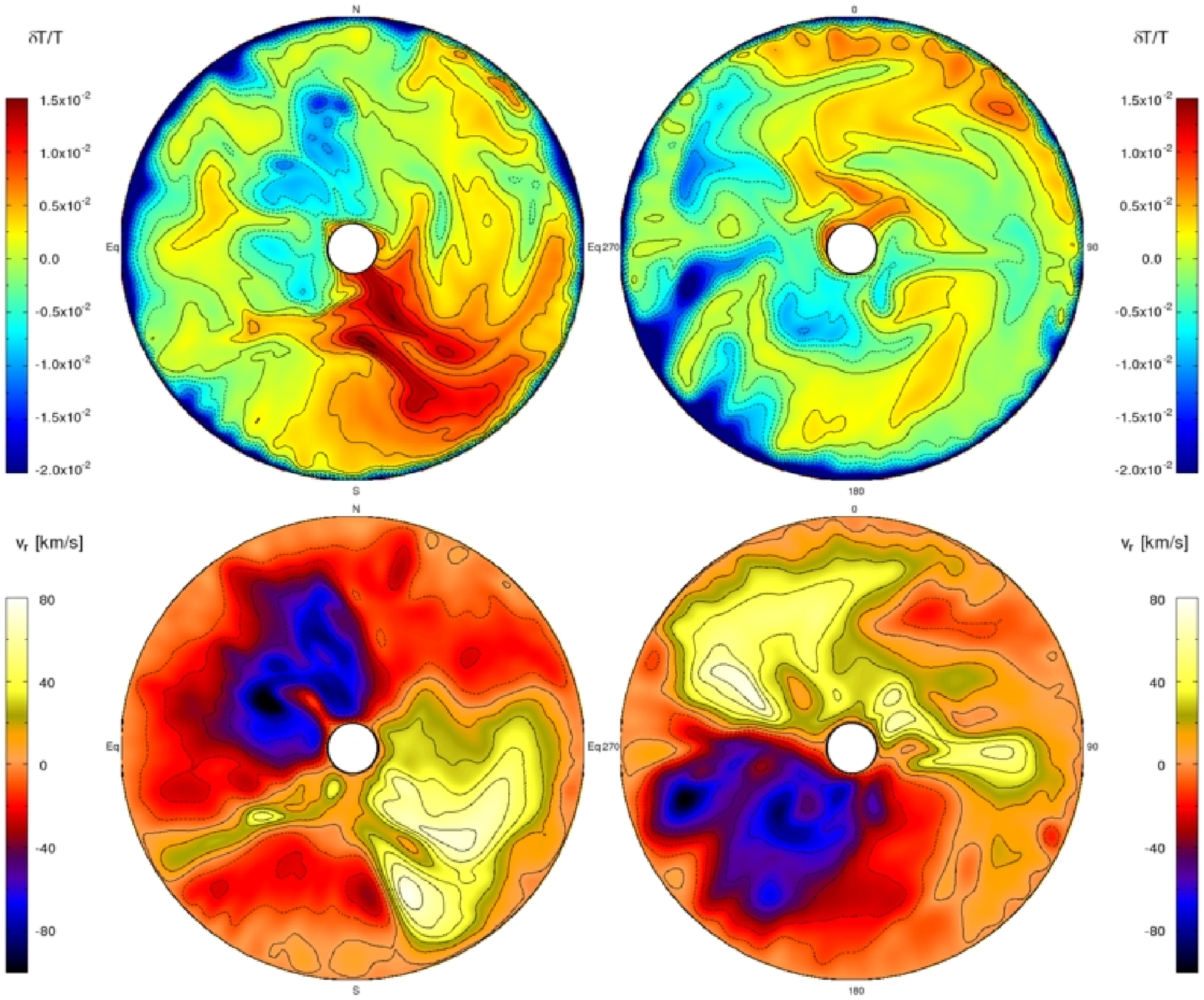}
\caption{
Slices of $\dT$ (top) and $v_r$ (bottom), in meridional (left) and
equatorial (right) planes in the non-rotating $T_{\rm c,8}=7$
simulation. Solid contours denote positive, dashed ones negative
values.
}
\lFig{slices}
\end{figure*}

In the non-rotating models there is no preferred axis in the
calculation. The expansion in spherical harmonics might be expected to
produce an artificial preferred axis along $\theta=0$, but the observed
dipole is not aligned with this axis. Neither is it aligned with any
perturbation in the initial conditions. The only explanation is some
form of ``spontaneous symmetry-breaking'', growing from numerical
noise in the calculation.

Admittedly, we cannot address precisely the nature of the flow at the
center with our method. As explained in \Section{models}, we solve the
anelastic hydrodynamics equations subject to a solid inner boundary
condition, which we have placed at $R=50$ km.  The sharp drop in the
radial velocity component at $R<100$ km/s is artificial, caused by the
solid inner boundary condition. The actual radial velocity component
at the center can be obtained by extrapolation of its value at 100 km
to the origin, as suggested by the large r.m.s. angular velocities
near the center. This implies that the center of the star is not at a
calm region at all, as one might expect from one-dimensional
calculations or multidimensional calculations that do not carry a full
360 degrees. Similar conclusion were reached by a recent 2D
anelastic study of convection in the interior of giant gas planets
\citep{Evo05}.

\subsection{The Temperature Fluctuation Spectrum}
\lSect{spectrum}

\begin{figure*}[htp]
\includegraphics[width=\textwidth]{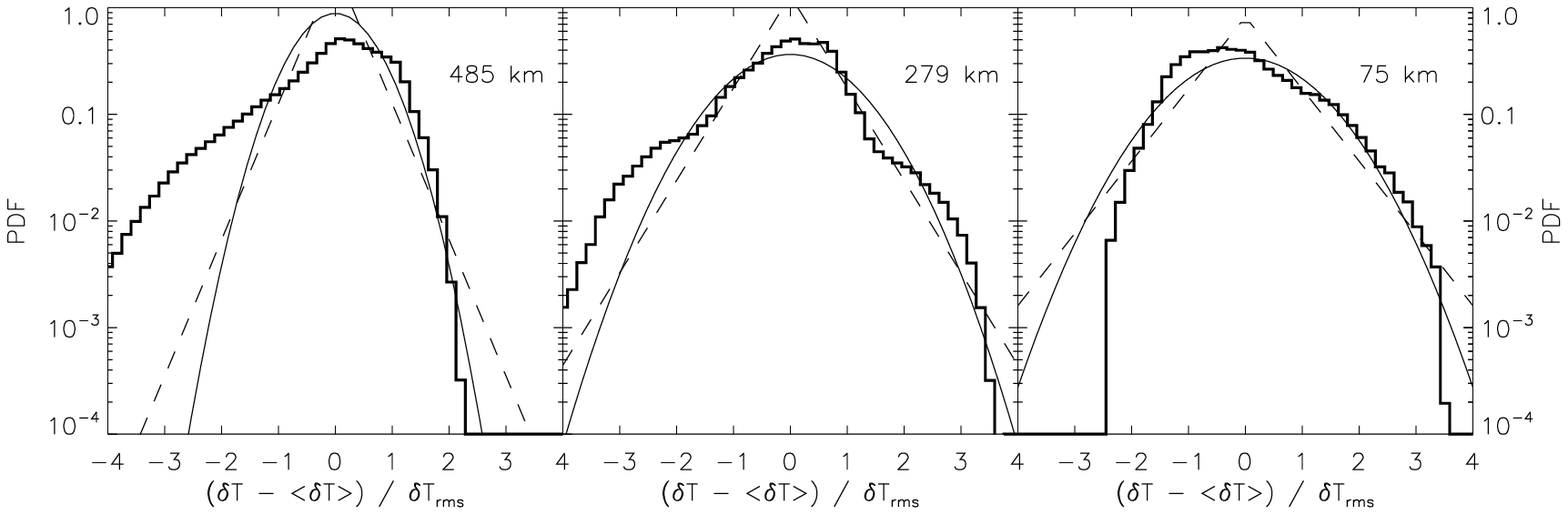}
\caption{
The probability distribution of temperature fluctuation evaluated in
three constant radius shells (485, 279, and 75 km), when the central
temperature is $T_{c,8}$ = 7.0. The y-axis shows the fraction of the
area in this shell that intersects fluid elements with temperatures
deviating by the amount shown from the mean, divided by the
root-mean-square of the temperature fluctuations in that shell. The
thin solid and dashed lines show best-fit Gaussian and Exponential
PDF, respectively. Only positive fluctuations were used in the fit,
and the PDF were constrained to be centered at zero.
}
\lFig{Tdist}
\end{figure*}

As discussed in \citet{Woo03}, the nature of the probability
density function (PDF) of temperature fluctuations is 
important for determining whether the supernova explosion will have
one or multiple ignition points. The sharper the high temperature tail
of the PDF, the more material there is with temperature just a bit
less than the ignition temperature of the first flame.


We have numerically estimated the temperature fluctuation PDF in our
simulation for constant radius shells of thickness $\delta R \simeq 2$
km. In each shell we subtract from every temperature fluctuation the
mean $\langle\dT\rangle$, divide by the r.m.s. $\dT_{\rm rms}$, and
use the resulting quantity as our independent variable: $(\dT -
\langle\dT\rangle)/\dT_{\rm rms}$. We then determine the PDF by
calculating the fraction of volume of the shell occupied by a given
fluctuation and dividing by the total volume of the shell. The
resulting PDF at three radii (485, 279, and 75 km), corresponding to
20 km from the top, the center of the modeled region, and 20 km
from the bottom, are shown as histograms in \Figure{Tdist}. We have
found best-fitting Gaussian distribution (GPDF) and exponential
distribution (EPDF) and plotted them as thin solid and dashed lines,
respectively. The PDF were fit via simple $\chi^2$ minimization,
constrained to be centered at $\dT = \langle\dT\rangle$, and only the
positive fluctuations were used in the fit.

In all three locations the GPDF produced a better fit than the
EPDF. None of the fits match the distribution perfectly over the
entire range of positive fluctuations, but in particular in the
$R=75$km shell the Gaussian fit follows the distribution out to $3.5
\dT_{\rm rms}$. The measured distributions at $R=485$ and 279 km
exhibit a strong excess of negative temperature fluctuations over
either GPDF or EPDF. This excess is entirely composed of downwards
flowing material, making up the \textit{return current} of the global
dipole flow described in \Section{dipole}.

Niemela \etal (2000) experimentally determined the temperature
fluctuation PDF in an incompressible fluid (cryogenic helium at 4 K)
as a function of Rayleigh number. At Rayleigh numbers comparable to
the ones we have simulated in this study, they also found a Gaussian
PDF.  At higher Rayleigh number, however, they observed a transition
to an Exponential PDF, which remained intact all the way up to Ra
$\sim 10^{15}$, the limit of their experiment. Even this is many
orders of magnitude below the Ra $\sim 10^{25}$ expected in the
centers of white dwarfs, and for all practical purposes the true
nature of the temperature fluctuation PDF before the onset of
explosive carbon burning must be considered unknown. Further numerical
studies at higher Rayleigh numbers are necessary, and may well detect
a transition to an exponential PDF.

\subsection{The Size and Persistence of Temperature Fluctuations}
\lSect{persistence}

While the PDF captures how rare a temperature fluctuation is,
averaged over space and time, it contains no information about the size
and duration of these fluctuations. The persistence of a fluctuation
being carried outward is of crucial importance in determining how far
from the center the hot spot will run away and ignite the supernova
explosion. A hot spot of a given size will be shredded to pieces on a
timescale set by the degree of turbulence within the outflow. Together
with the laminar outflow velocity, this time scale sets the maximum
distance from the center that a hot spot can reach. As we discussed in
\Section{vel+pert}, we see in our simulations the beginning of a
turbulent cascade. Unfortunately, our resolution is nowhere near
adequate to follow this cascade from the integral length scale (the
size of the largest eddies) all the way down to the Kolmogorov scale
(the smallest eddies). We can, however, make use of scaling relations
to infer the minimum size fluctuations that will survive out to a
given distance from the center.

As discussed by \citet{Woo03}, the spots within the dipole flow that
runaway first will be those in which heating by burning just
compensates cooling by adiabatic expansion. This consideration, plus a
dipole flow speed of approximately 50 - 100 km s$^{-1}$, naturally
gives ignition about 100 km out from the center of the star. 

Within the dipole flow we find a residual random velocity field. The
smallest resolved eddies have a length scale of about 50 km and a
typical velocity of about 10 km s$^{-1}$. Assuming Kolmogorov-Obukhov
scaling ($v \propto L^{1/3}$) we can extrapolate to below our
resolution limit. We estimate that an eddy 1 km in size will turn over
in about 1 second. Smaller eddies will turn over faster, so we take
this as a characteristic volume for the fluid that will mix during the
1 to 2 seconds it takes a hot perturbation to move to the ignition
radius -- 100 km.


All these numbers will require a much better resolved calculation to
be taken precisely, but they suggest that the perturbations that
ignite the runaway will be moderately large, $\sim1$ km, a value that
might even be resolved in future numerical studies.

Our results also show significant correlation between hot spots. They
are not scattered randomly throughout the dipole flow, but
concentrated near its axis. Temperature differences will be amplified
in the subsequent runaway by the high power of the reaction rate
temperature sensitivity, but a reasonable picture of the ignition
might be a single, broad, off-center ``plume'' rather that multiple
hot spots.

\subsection{The Effect Of Rotation}
\lSect{rotation}

In addition to the non-rotating models discussed in the previous
sections, we also calculated one model (C7rot) with a rigidly rotating
background state. Rotation is introduced into the simulation by
solving the anelastic conservation equations in a rotating reference
frame. This results in the Coriolis force term on the right hand side
of Eq.(2b) of $2\rhobar\;\mathbf{v}\times\Omegabar$, where $\Omegabar$
is the reference state rotation rate vector in rad/sec. The
centrifugal term ($\rhobar\;\Omegabar\times\Omegabar\times\mathbf{r}$)
is neglected, because it is usually much smaller than the
gravitational force.\footnote{The centrifugal term must be included
for situations where the centrifugal and gravitational forces are
comparable.}

\begin{figure}[htp]
\includegraphics[width=\columnwidth]{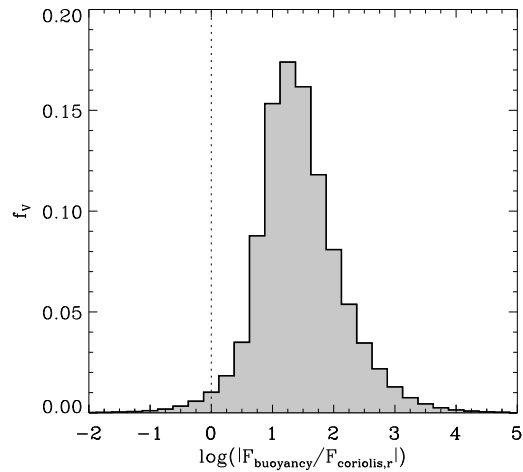}
\caption{
Distribution (volume fraction) of the ratio of the buoyancy force
($-\mathbf{g}\;\drho/\rhobar$) to the radial component of
the Coriolis force ($2\;(\mathbf{v} \times \Omegabar)_r$). Material
to the left of the dotted vertical line experiences a Coriolis force
that is stronger than the local buoyancy force.
}
\lFig{coriolis}
\end{figure}

\begin{figure*}[htp]
\includegraphics[width=\textwidth]{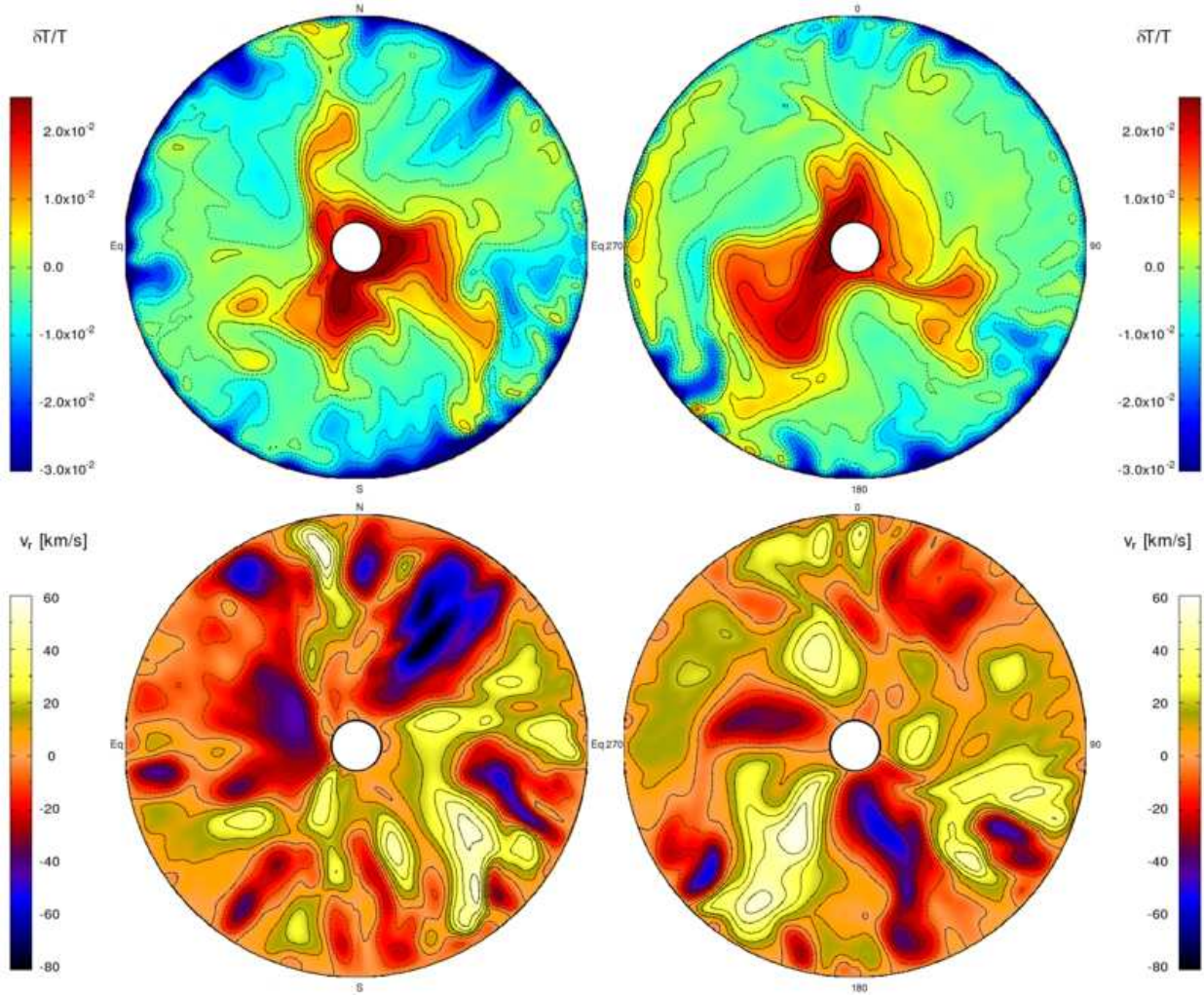}
\caption{
Like \Figure{slices}, but for the C7rot simulation. A rotating
background state seems to break up the dipole flow.
}
\lFig{slices_rot}
\end{figure*}

Accreting white dwarfs are thought to rotate quite rapidly, with
rotation rates approaching Keplerian at the surface
\citep[eg.][]{Yoo04}.  In the \Kepler\ model which is the basis for
this anelastic study, the Keplerian rotation rate at the surface is
$\Omega_0 = (GM/R^3)^{1/2} = 5.6$ rad/sec. We are interested in
determining what effect even a small amount of rotation has on the
dipole flow that we observed in the non-rotating simulations. For this
purpose, we chose a moderate background rotation rate of 0.167
rad/sec, only 3\% of $\Omega_0$. As we shall see, even this
comparatively small amount of rotation has an appreciable effect upon
the fluid flow. At the outer boundary of our simulation the ratio of
centrifugal to gravitational force is $\bar{\Omega}^2 R^3/(G M) =
7\times10^{-5}$, justifying the omission of the centrifugal force
term. The importance of the coriolis force relative to buoyancy forces
is addressed in \Figure{coriolis}. We have plotted the fraction of the
total simulated volume at a given ratio of buoyancy force
($-\bar{\mathbf{g}} \drho/\rhobar$) to the radial component of
coriolis force versus the logarithm of this ratio. This plot shows
that buoyancy dominates coriolis forces for most of the simulated
volume: only for 1.3\% of the total volume does the radial component
of the coriolis force exceed the buoyancy force. Two further
dimensionless numbers characterize the importance of rotation. The
Ekman number is a measure of the relative importance of viscous to
Coriolis, and the Rossby number of inertial to rotational forces.
\begin{subequations}
\be
{\rm Ek} \sim \f{\nubar}{2\Omegabar D^2}
\ee
\be
{\rm Ro} \sim \f{v}{2D\Omega\sin{\theta}}, 
\ee
\end{subequations} 
where $\theta$ is the co-latitude. We have Ek $\approx 7 \times
10^{-4}$, and at the equator Ro=1 or 1/3, depending on whether we use
$v_{\rm max}$ or $v_{\rm mean}$.

Even though the centrifugal and coriolis forces are small compared
with gravity and buoyancy, respectively, the rotating background state
significantly alters the flow in the simulation. \Figure{slices_rot}
shows equatorial and meridional slices of temperature perturbation and
radial velocity for the C7rot simulation. A comparison with
\Figure{slices} shows that the dipole flow has pretty much
disappeared. Instead of a well defined outflow (inflow) in the lower
right (upper left) quadrant of the meridional slice in the C7
simulation, both inflowing and outflowing plumes can be found in all
four quadrants of the same slice in the C7rot simulation. The
temperature fluctuations in C7rot are also broken up into multiple hot
and cold spots, instead of one hot and one cold plume in
C7. Two-dimensional simulations at a much higher Rayleigh number
show a transition from a dipole to a differentially rotating
longitudinal flow as the rotation rate is increased \citep{Evo05}. It
is important to note, however, that flow through the center might
still be possible in a different configuration. The temperature
fluctuations and convective velocities are of the same order as in the
C7 simulation, and the results of \Section{spectrum} and
\ref{sec:persistence} will continue to hold as long as there is at
least one central outflow of hot material. An asymmetric supernova
explosion, however, becomes less likely in the absence of a strong
dipole flow.

\section{CONCLUSIONS}
\lSect{conclusion}

Three-dimensional studies at moderate Rayleigh number ($\sim10^7$) of
carbon ignition in the highly degenerate core of a white dwarf star
show a dipole nature to the flow. If such a flow pattern persists at
the much higher Rayleigh number of the actual white dwarf, this may
lead to the asymmetric ignition and explosion of Type Ia supernovae.
We also find that a moderate degree of rotation, corresponding to less
than 10\% critical at the surface, disrupts this dipole, leading to
ignition over a broader range of angles.

Within the flow, there still exists turbulence and this will limit the 
persistence of temperature fluctuations that might serve as ignition 
points. Only the larger fluctuations will survive the transit through
the core to the estimated ignition radius of 100 km. We estimate those
fluctuations will be a km or larger. 

Finally we have estimated the PDF for the temperature fluctuations
(Gaussian) and commented on the degree of spatial correlation for the
hot spots (high). This suggests that once a runaway ignites in some
locality, other points will swiftly ignite in approximately the same
vicinity (although the ignition may continue for some time).

Though the present study has been the first to follow the burning in
3D in a large fraction of the unstable core, it has raised many
questions that will require further study.

Chief among them are the scaling properties of the dipole flow with
Reynolds number. The present study has a Reynolds number approximately
10$^{11}$ times smaller than the actual star. It is quite likely that the
actual flow becomes increasingly chaotic at larger Reynolds
number, analogous to what is seen in Rayleigh-Bernard convection
\citep{Kad01}. Two-dimensional simulations are capable of showing
the dipole flow when run for full cylinders, and might have adequate
resolution to study this scaling.

Second, though we have argued that its effects are minimal, the
artificial inner boundary condition in the present study is
troublesome and needs to be removed. This is not an inherent
deficiency in the anelastic method, but results from the use of a
spectral method in spherical coordinates -- the angular derivatives
diverge at $r=0$. A Cartesian, finite-volume version of the code has
been developed by \citet{Evo05} and is currently being used to further
study the present problem. Initial 2D simulations of convection in
giant gaseous planets with this code show that the presence of a solid
core in the non-rotating case can lead to a significantly different
flow structure. With increasing rotation rate, however, the flow
settles into a differentially rotating profile that is only weakly
affected by the presence of a solid core. Thus it may not be necessary
to remove the solid inner core for simulations of rapidly rotating
white dwarfs.

Third, the study needs to be started earlier (at a lower central
temperature) and run longer (very many convective turnover times). It
takes time for the 1D background state model to adjust to the new
code. Ideally, one wants to watch the gradual rise in the overall
temperature until the runaway actually ignites in localized
regions. One can do that by gradually adjusting the background state
in the anelastic model. Here we sought stability at the expense of
imposing an artificially large radiative boundary at the outer edge of
the problem.

Given the importance of the ``ignition problem'' to understanding Type
Ia supernovae, we expect significant progress on all these fronts in
the near future.

We thank M. Zingale for informative
discussions. \Figure{dipole_flow_3D} was created with IFrIT, a
visualization tool written by Nick Gnedin.  This work has been
supported by the NSF (AST 02-06111), NASA (NAG5-12036), and the DOE
Program for Scientific Discovery through Advanced Computing (SciDAC;
DE-FC02-01ER41176. All computations were performed on NERSC's
\textit{Seaborg}, a 6,080 processor IBM SP RS/6000 supercomputer.


\begin{thebibliography}{99}

\bibitem[Baraffe et al.(2004)]{Bar04} 
Baraffe, I., Heger, A., \& Woosley, S.~E.\ 2004, \apj, 615, 378

\bibitem[Boyd (2000)]{Boy00}
Boyd, J.~P.\ 2000, Chebyshev and Fourier Spectral Methods (2nd ed.; Mineola, NY: Dover)

\bibitem[Evonuk \& Glatzmaier (2005)]{Evo05}
Evonuk, M., \& Glatzmaier, G.~A. 2005, Planetary and Space Science, submitted

\bibitem[Gamezo et al.(2003)]{Gam03} 
Gamezo, V.~N., Khokhlov, A.~M., Oran, E.~S., Chtchelkanova, A.~Y., \&
Rosenberg, R.~O.\ 2003, Science, 299, 77

\bibitem[Garcia-Senz \& Woosley(1995)]{Gar95} 
Garcia-Senz, D., \& Woosley, S.~E.\ 1995, \apj, 454, 895

\bibitem[Garc{\'{\i}}a-Senz \& Bravo(2005)]{Gar05} 
Garc{\'{\i}}a-Senz, D., \& Bravo, E.\ 2005, \aap, 430, 585 
 
\bibitem[Glatzmaier(1984)]{Gla84} 
Glatzmaier, G. A. 1984, J. Comp. Phys., 55, 461 

\bibitem[Glatzmaier \& Roberts(1995)]{Gla95} 
Glatzmaier, G. A., \& Roberts, P.~H. 1995, Nature, 377, 203

\bibitem[H\"oflich \& Stein(2001)]{Hof01}
H\"oflich, P., \& Stein, J. 2002, \apj, 568 779

\bibitem[Hillebrandt \& Niemeyer(1999)]{Hil99} 
Hillebrandt, W., \& Niemeyer J. 2000, ARAA, 38, 191b 

\bibitem[Kadanoff(2001)]{Kad01}
Kadanoff, L. P. 2001, Physics Today, 54(8), 34

\bibitem[Kraichnan(1962)]{Kra62}
Kraichnan, R. H. 1962, Phys. of Fluids, 5, 1374

\bibitem[Lantz \& Fan (1999)]{Lan99}
Lantz, S. R., \& Fan, Y. 1999, \apj, 121, 247.

\bibitem[Niemela et al.(2000)]{Niem00}
Niemela, J. J., Skrbek, L., Sreenivasan, K. R., \& Donnelly,
R. J. 2000, Nature, 404, 837

\bibitem[Niemeyer \& Hillebrandt(1995)]{Nie95} 
Niemeyer, J.~C., \& Hillebrandt, W.\ 1995, \apj, 452, 769

\bibitem[Niemeyer, Hillebrandt, \& Woosley(1996)]{Nie96} 
Niemeyer, J. C., Hillebrandt, W., \& Woosley, S. E. 1996, \apj, 471, 903

\bibitem[Niemeyer \& Woosley(1997)]{Nie97} 
Niemeyer, J. C., \& Woosley, S. E. 1997, \apj, 475, 740

\bibitem[Niemeyer(1999)]{Nie99}
Niemeyer, J. C. 1999, ApJL, 523, L57

\bibitem[Hillebrandt \& Niemeyer(2000)]{Nie00} 
Hillebrandt, W., \& Niemeyer, J.~C.\ 2000, \araa, 38, 191

\bibitem[Nomoto, Sugimoto, \& Thielemann(1976)]{Nom76} 
Nomoto, K., Sugimoto, D., \& Neo, S. 1976,  Ap\&SS, 39L, 37

\bibitem[Nomoto, Thielemann, \& Yokoi(1984)]{Nom84} 
Nomoto, K., Thielemann, F.-K., \& Yokoi, K. 1984, \apj, 286, 644

\bibitem[Plewa et al.(2004)]{Ple04} 
Plewa, T., Calder, A.~C., \& Lamb, D.~Q.\ 2004, \apjl, 612, L37

\bibitem[Porter \& Woodward(2000)]{Por00}
Porter, D. H., \& Woodward, P. R. 2000, ApJS, 127, 159

\bibitem[R{\" o}pke \& Hillebrandt(2005)]{Roe05a} 
R{\" o}pke, F.~K., \& Hillebrandt, W.\ 2005, \aap, 431, 635
 
\bibitem[R{\" o}pke et al. (2005)]{Roe05b} 
R{\" o}pke, F.~K., Hillebrandt, W., Niemeyer, J. C., \& Woosley, S. E.
2005, \aap, submitted

\bibitem[Sun, Schubert, \& Glatzmaier(1993)]{Sun93}
Sun, Z.-P., Schubert, G., \& Glatzmaier, G.~A. 1993, Science, 260, 661
 
\bibitem[Timmes \& Swesty(2000)]{Tim00}
Timmes, F.X., \& Swesty, F.D., 2000, \apj\ Supplement Series, 126, 501

\bibitem[Weaver, Zimmerman, \& Woosley(1978)]{Wea78} 
Weaver, T. A., Woosley, S. E., \& Zimmerman, G. B. 1978, \apj, 225, 1021

\bibitem[Woodward et al.(2003)]{Wood03} 
Woodward, P.~R., Porter, D.~H., \& Jacobs, M.\ 2003, ASP
Conf.~Ser.~293: 3D Stellar Evolution, 293, 45. see also
www.lcse.umn.edu/3Dstars.

\bibitem[Woosley, Wunsch, \& Kuhlen(2003)]{Woo03}
Woosley, S. E., Wunsch, S., \& Kuhlen, M. 2003, \apj, 607, 921 

\bibitem[Wunsch \& Woosley(2004)]{Wun04} 
Wunsch, S., \& Woosley, S.~E.\ 2004, \apj, 616, 1102
 
\bibitem[Yoon \& Langer(2004)]{Yoo04} 
Yoon, S.-C., \& Langer, N.\ 2004, \aap, 419, 623

\end{thebibliography}
\end{document}